\documentclass[10pt,preprint2]{aastex}
\begin{document}
\def \Angs        {\hbox{\rm \AA}}
\def \Vvec        {\hbox{$\mathbf V$}}
\def \nvec        {\hbox{$\mathbf {\hat e}_n$}}
\def \zvec        {\hbox{$\mathbf {\hat e}_z$}}
\def \OTSH        {\hbox{\rm OTSh}}
\def \VW          {\hbox{$V_W$}}
\def \VWpara      {\hbox{$V_{W,\parallel}$}}
\def \VWperp      {\hbox{$V_{W,\perp}$}}
\def \VP          {\hbox{$V_P$}}
\def \V8          {\hbox{$V_8$}}
\def \Vrel          {\hbox{$V_{\rm rel}$}}
\def \TA          {\hbox{$T_{\rm A}$}}
\def \NNc          {\hbox{$\cal N$}}
\def \Tmax          {\hbox{$T_{\rm Max}$}}
\def \EMA          {\hbox{$EM_{\rm A}$}}
\def \HH          {\hbox{$\cal H$}}
\def \Rbase          {\hbox{$R_{\rm base}$}}
\def \Rclump          {\hbox{$R_{\rm clump}$}}
\def \jem          {\hbox{$j_{\rm em}$}}
\def \VPpara      {\hbox{$V_{P,\parallel}$}}
\def \VPperp      {\hbox{$V_{P,\perp}$}}
\def \Ain      {\hbox{$A_{1}$}}
\def \Aex      {\hbox{$A_{2}$}}
\def \DL       {\hbox{$\Delta \ell$}} 

\def \lam      {\hbox{$\lambda$}}
\def \Llam      {\hbox{$L_\lambda$}}
\def \wlam      {\hbox{$w_\lambda$}}
\def \kaplam    {\hbox{$\kappa_\lambda$}}
\def \Tw      {\hbox{$T_w$}}
\def \gamav   {\hbox{$\gamma_{\rm v}$}}
\def \gamaF   {\hbox{$\gamma_F$}}
\def \Rc   {\hbox{$R_c$}}
\def \Dclump   {\hbox{$D_c$}}
\def \dellam  {\hbox{$\delta \lambda$}}
\def \delw    {\hbox{$\delta w$}}
\def \flam    {\hbox{$f_\lambda$}}
\def \Flam    {\hbox{$F_\lambda$}}
\def \lamflam {\hbox{$\lambda f_\lambda$}}
\def \Lamlam {\hbox{$\Lambda_{\lambda}$}}
\def \gammaf  {\hbox{$\gamma_F$}}
\def \gammaw  {\hbox{$\gamma_W$}}
\def \taulam    {\hbox{$\tau_\lambda$}}
\def \tauff    {\hbox{$\tau_{\rm ff}$}}
\def \HeII{\hbox{He~{\sc ii}}}
\def \Stromgren       {\hbox{Str\"{o}mgren}}
\newcommand{\Halpha}{\mbox{H$\alpha$}}
\newcommand{\kmsec}{\mbox{$\rm{km \; s^{-1}}$}}
\newcommand{\Lnu}{\mbox{$L_\nu$}}
\newcommand{\EM}{\mbox{$EM$}}
\newcommand{\EMT}{\mbox{$EM(T)$}}
\newcommand{\Lstar}{\mbox{$L_*$}}
\newcommand{\jnu}{\mbox{$j_\nu$}}
\newcommand{\mH}{\mbox{$m_{\rm H}$}}
\newcommand{\mstar}{\mbox{$M_*$}}
\newcommand{\msun}{\mbox{$M_\odot$}}
\newcommand\Rsun{\hbox{$R_\odot$}}
\newcommand\Lsun{\hbox{$L_\odot$}}
\newcommand\Msun{\hbox{$M_\odot$}}
\newcommand\Rstar{\hbox{$R_*$}}
\newcommand\Tstar{\hbox{$T_*$}}
\newcommand\Teff{\hbox{$T_{eff}$}}
\newcommand\TL{\hbox{$T_L$}}
\newcommand\TX{\hbox{$T_X$}}
\newcommand\THHe{\hbox{$T_{HHe}$}}
\newcommand\TGR{\hbox{$T_{GR}$}}
\newcommand\hhe{\hbox{$H-He$}}
\newcommand\HWHM{\hbox{$HWHM$}}
\newcommand\EMX{\hbox{$EM_{\rm X}$}}
\newcommand\fir{\hbox{$fir$}}
\newcommand\eg{\hbox{e.g.,}}
\newcommand\ie{\hbox{i.e.,}}
\newcommand\etal{\hbox{et~al.}}
\newcommand\ROSAT{\hbox{\it ROSAT}}
\newcommand\ASCA{\hbox{\it ASCA}}
\newcommand\XMM{\hbox{\it XMM-Newton}}
\newcommand\Chandra{\hbox{\it Chandra}}
\newcommand\Copernicus{\hbox{\it Copernicus}}
\newcommand\Einstein{\hbox{\it Einstein}}
\newcommand\EUVE{\hbox{\it EUVE}}
\newcommand\Rlow{\hbox{$R_{\rm low}$}}
\newcommand\Rmin{\hbox{$R_{\rm min}$}}

\newcommand\CIV{\hbox{C {\sc iv}}}
\newcommand\NVII{\hbox{N {\sc vii}}}
\newcommand\NVI{\hbox{N {\sc vi}}}
\newcommand\NV{\hbox{N {\sc v}}}
\newcommand\OVI{\hbox{O {\sc vi}}}
\newcommand\OVII{\hbox{O {\sc vii}}}
\newcommand\OVIII{\hbox{O {\sc viii}}}
\newcommand\MgXI{\hbox{Mg {\sc xi}}}
\newcommand\MgXII{\hbox{Mg {\sc xii}}}
\newcommand\FeXVII{\hbox{Fe {\sc xvii}}}
\newcommand\NeIX{\hbox{Ne {\sc ix}}}
\newcommand\NeX{\hbox{Ne {\sc x}}}
\newcommand\SiIV{\hbox{Si {\sc iv}}}
\newcommand\SiXIII{\hbox{Si {\sc xiii}}}
\newcommand\SiXIV{\hbox{Si {\sc xiv}}}
\newcommand\SXV{\hbox{S {\sc xv}}}
\newcommand\SXVI{\hbox{S {\sc xvi}}}
\newcommand\ArXVII{\hbox{Ar {\sc xvii}}}
\newcommand\CaXIX{\hbox{Ca {\sc xix}}}
\newcommand\CaXX{\hbox{Ca {\sc xx}}}
\newcommand\CaXXI{\hbox{Ca {\sc xxi}}}
\newcommand\FeXX{\hbox{Fe {\sc xx}}}
\newcommand\FeXXII{\hbox{Fe {\sc xxii}}}
\newcommand\FeXXIII{\hbox{Fe {\sc xxiii}}}
\newcommand\FeXXIV{\hbox{Fe {\sc xxiv}}}
\newcommand\FeXXV{\hbox{Fe {\sc xxv}}}
\newcommand\ftoi{\hbox{$f/i$}}
\newcommand\PV{\hbox{P {\sc v}}}
\newcommand\HtoHe{\hbox{$H/He$}}
\newcommand\RRR{\hbox{\sf R}}
\newcommand\zpup{\hbox{$\zeta\ ${\rm Pup}}}
\newcommand\zori{\hbox{$\zeta\ ${\rm Ori}}}
\newcommand\zoriA{\hbox{$\zeta\ ${\rm Ori A}}}
\newcommand\COB{\hbox{Cyg OB2}}
\newcommand\cygate{\hbox{Cyg OB2 No. 8a}}
\newcommand\cygnin{\hbox{Cyg OB2 No. 9}}
\newcommand\dori{\hbox{$\delta\ ${\rm Ori}}}
\newcommand\doriA{\hbox{$\delta\ ${\rm Ori A}}}
\newcommand\tsco{\hbox{$\tau\ ${\rm Sco}}}
\newcommand\thoriC{\hbox{$\theta^1$ Ori C}}
\newcommand\eori{\hbox{$\epsilon\ ${\rm Ori}}}
\newcommand\ecma{\hbox{$\epsilon\ ${\rm CMa}}}
\newcommand\bcma{\hbox{$\beta\ ${\rm CMa}}}
\newcommand\sori{\hbox{$\sigma\ ${\rm Ori}}}
\newcommand\iori{\hbox{$\iota\ ${\rm Ori}}}
\newcommand\zoph{\hbox{$\zeta\ ${\rm Oph}}}
\newcommand\gcas{\hbox{$\gamma\ ${\rm Cas}}}
\newcommand\bcru{\hbox{$\beta\ ${\rm Cru}}}
\newcommand\xper{\hbox{$\xi\ ${\rm Per}}}
\newcommand\Berghofer{\hbox{Bergh\"{o}fer}}
\newcommand\vinf{\hbox{$v_\infty$}}
\newcommand\Mdot{{\hbox{$\dot M$}}}
\newcommand\Ndot{{\hbox{$\dot N$}}}
\newcommand\Msunyr{\hbox{$M_\odot\,$yr$^{-1}$}}
\newcommand\kms{\hbox{km$\,$s$^{-1}$}}
\newcommand\LXLBol{{\hbox{$L_X/L_{Bol}$}}}
\newcommand\Ne{\hbox{$N_e$}}
\newcommand\Np{\hbox{$N_p$}}
\newcommand\lamz{\hbox{$\lambda_{\circ}$}}
\newcommand\linflux{\hbox{line flux$/10^{-13}$}}
\newcommand\censhif{\hbox{$\frac{\displaystyle{\rm
centroid}}{\displaystyle{\rm shift}}$}}
\newcommand\shoceff{\hbox{$\frac{\displaystyle{\rm shock}}{\displaystyle{\rm
efficiency}}$}}
\newcommand{\n}[1]{\tablenotemark{\#1}}
\newcommand{\vsini}{$v \sin i$}
\newcommand\Rfir{\hbox{$R_{fir}$}}
\newcommand\Rtau{\hbox{$R_{\tau = 1}$}}
\newcommand\LB{{\it line-blending effects}}

%\begin{document}

%\slugcomment{Submitted to {\it The Astrophysical Journal}: xxxxx, 2007}
\shorttitle{X-ray Line Emission from OB wind clumps}

\title{The Effects of Clumps in Explaining X-ray Emission Lines from 
Hot Stars}

%\author {J.~P. Cassinelli$^{1}$, R. Ignace$^{2}$,W.~L. Waldron$^{3}$,
%J. Cho$^{4}$, N. A. Murphy$^{1}$, and A. Lazarian$^{1}$\\
%$^{1}$Department of Astronomy, University of Wisconsin-Madison, Madison, WI 53711,
%cassinelli@astro.wisc.edu; \\
%$^{2}$Department of Physics, Astronomy, \& Geology,
%East Tennessee State University, Johnson City, TN, 37614, ignace@etsu.edu; \\
%$^{3}$Eureka Scientific Inc., 2452 Delmer St., Oakland, CA, 94602, 
%wwaldron@satx.rr.com; \\ $^{4}$ Department of Astronomy and Space
%Science, Chungnam National University, 
%Daejeon, Korea, jcho@cnu.ac.kr}

\author {J.~P.~Cassinelli,}
\affil{
Department of Astronomy, University of Wisconsin-Madison, Madison, WI 53711,\\
cassinelli@astro.wisc.edu}

\author{R.~Ignace,}
\affil{
Department of Physics, Astronomy, \& Geology,
East Tennessee State University, Johnson City, TN, 37614, \\ ignace@etsu.edu}

\author{W.~L.~Waldron,}
\affil{
Eureka Scientific Inc., 2452 Delmer St., Oakland, CA, 94602,  \\
wwaldron@satx.rr.com}

\author{J.~Cho,}
\affil{
Department of Astronomy and Space
Science, Chungnam National University, 
Daejeon, Korea, \\ jcho@cnu.ac.kr}

\author{N.~A.~Murphy, and A.~Lazarian}
\affil{
Department of Astronomy, University of Wisconsin-Madison, Madison, WI 53711, \\
murphy@astro.wisc.edu, lazarian@astro.wisc.edu
}

\begin{abstract}

It is now well established that stellar winds of hot stars are fragmentary
and that the X-ray emission from stellar winds has a strong contribution
from shocks in winds. \Chandra\ high spectral resolution observations of
line profiles of O and B stars have shown numerous properties that had
not been expected. Here we suggest explanations by considering the X-rays
as arising from bow shocks that occur where the stellar wind impacts on
spherical clumps in the winds. We use an accurate and stable numerical
hydrodynamical code to obtain steady-state physical conditions for the
temperature and density structure in a bow shock.  We use these solutions
plus analytic approximations to interpret some major X-ray features:
the simple power-law distribution of the observed emission measure
derived from many hot star X-ray spectra and the wide range of ionization
stages that appear to be present in X-ray sources throughout the winds.
Also associated with the adiabatic cooling of the gas around a clump is
a significant transverse velocity for the hot plasma flow around the clumps,
and this can help to understand anomalies associated with observed line
widths, and the differences in widths seen in stars with high and low
mass-loss rates. The differences between bow shocks and the planar shocks
that are often used for hot stars are discussed. We introduce an ``on
the shock'' (\OTSH) approximation that is useful for interpreting the
X-rays and the consequences of clumps in hot star winds and elsewhere
in astronomy.  

\end{abstract}


\section{Introduction} 

Our goal is to explain with clump bow shocks many of the problems found
in the survey of 17 normal OB stars by Waldron \& Cassinelli (2007)
(hereafter WC07). Well-resolved spectral line profiles are found to be
neither shifted nor skewed to the blue (i.e., shortward) as had been
expected (e.g., MacFarlane \etal\ 1991).  The line widths are broader for
supergiants and stars with thick winds than for lower luminosity stars.
It seems clear that X-rays are formed in numerous shock fragments
distributed throughout the wind. 

The topic of high non-radiative equilibrium temperatures in the winds of 
hot stars began with the discovery of superionization stages seen in the FUV
spectra obtained with the \Copernicus\ Satellite. Lamers \& Morton (1976)
analyzed the spectrum of \zpup\ (O4If) and found strong lines of \OVI\
(1040~\Angs) and \NV\ (1240~\Angs). These ionization stages seemed to
require that the winds are ``warm'' with temperatures of about $2 \times
10^5$~K. Alternatively, the ions could be produced by the Auger ionization
process whereby 2 electrons are removed from the dominant ionization stage
following K-shell ionization by X-rays (Cassinelli, Lamers, \& Castor 1978).
Cassinelli \& Olson (1979) used a ``thin corona plus cool wind'' model to
show that the X-rays from a spatially thin corona plus the Auger process
could explain observed boundaries in the HR~diagram: a sharp cutoff in the
presence of the \OVI\ UV line that occurs at spectral type B0, a cutoff of the
superionization line of \NV\ at B1.5, and similar ones for \CIV\ at B5 and
\SiIV\ at B8 (Cassinelli \& Abbott 1982). In each case the high ions could
be explained by the removal of 2 electrons from the {\it dominant} stage of
ionization, which are $O^{+3}$, $N^{+2}$, and $C^{+1}$. Odegard \& Cassinelli
(1982) explained the more complicated case of the \SiIV\ lines again using the
Auger process. Soon after the predictions based on the Auger mechanism were
made, X-rays were discovered from O stars in the first observations with the
\Einstein\ satellite (Seward \etal\ 1979; Harnden \etal\ 1979). However the
observed X-ray spectral distribution did not agree with the idea that the
X-rays were arising from a thin corona at the base of the cool wind, because
the attenuation of soft X-rays by the cool wind was absent. It became clear
that X-ray sources needed to occur farther out in the wind.  Lucy \& White
(1982) and Lucy (1982) proposed that the X-rays were generated from shocks
embedded in the wind and developed models of structured shocks. In fact, 
Lucy \& White (1982) proposed that the shocks were at the {\it outer} 
face of clumps being driven through the wind, and thus were bow shock
in nature.  
An advantage of the shock models is that they could form naturally by virtue
of the instability in the physics of line-driven winds (as first noted by
Lucy \& Solomon 1967).  Thus unlike the warm wind and the corona plus cool
wind models, shocks formed by line driven wind instabilities
do not need to assume an input of mechanical energy flux from the
star to heat either an extensive part of the wind nor a thin  coronal zone.

From a moderately high resolution spectrum of Orion belt stars with the
\Einstein\ satellite Solid State Spectrometer, Cassinelli \& Swank (1983)
found that X-ray line emission from \ion{Si}{13} and \ion{S}{15} in
\zori. These indicated the presence of gas that is hotter than had been 
needed to explain the softer X-ray flux. Since this relatively hard line 
radiation could escape from deep in the wind, they proposed that there could
still be zones, such as magnetic loops of very hot plasma, near the star's
surface. Because single OB stars are not noted for X-ray variability at more
than about the one percent level, Cassinelli \& Swank also concluded that
the embedded wind shocks could not be in the form of spherical shells, as
in the picture of Lucy (1982), but rather there must be of order $10^4$
distributed sources or ``shock fragments'' in the wind. The emission from a
number of sources at a range of heights in the wind would lead to a
statistically steady rate of X-ray production. Thus, even from relatively
early in the history of hot star X-ray astronomy it was thought that the
X-rays must arise from discrete distributed source regions, with the
possibility that hotter zones are present near the star. Further
improvements on that picture required the higher spectral resolution of
\Chandra\ and \XMM.

The nature of the line-driven wind instability that leads to shocked
regions was more fully explained by Owocki, Castor, \& Rybicki (1988),
who developed a radial, 1-D picture for the spatial distribution of
the shocks. Spherically symmetric shock models have also been computed,
for example by Cooper \&   Owocki (1990) and Feldmeier \etal\ (1997).
MacFarlane \& Cassinelli (1989) developed a basic model for an individual
shock as being in the form of a ``driven-wave'' with an inward boundary
on the star-ward side of the wave where the wind collides with the
driven wave, and an outward facing shock at the upper boundary where
the driven wave catches up with slower moving material ahead. The
shock properties were explained in terms of basic Rankine-Hugoniot
shock relations. The driven wave has a nearly constant pressure, set
by the dynamic pressure of the incident material ($\rho \Vrel^2$),
and the temperature of the gas can be quite high and is determined by
$\Vrel^2$, where \Vrel\ is the speed of the incident material relative
to the shock front. For strong shocks the density increases by a factor
of four relative to the incident material, and then in the layers of
the driven-wave where the gas radiatively cools back to the radiative
equilibrium value of the ambient wind, the density increases by about
3 orders of magnitude. It is likely that a shell of such a high density
contrast gas is unstable, such that the compressed gas will not remain
in the form of a spherically expanding shell, but instead break up into
clumps with densities of about $10^3$ times that of the ambient wind.  

The initiation of our  bow shock approach arose through our attempt to
understand the anomalous properties of the B0.5~V
star \tsco\ (Howk \etal\
2000). In \Copernicus\ spectra (Lamers \& Rogerson, 1978), the star shows
unusual red-shifted absorption features in the \OVI\ and \NV\ P-Cygni lines.
This could arise if there were infall of matter. In addition, \ROSAT\
observations of \tsco\ showed that the X-ray spectrum is unusually hard
relative to other early B~stars.  Howk \etal\ proposed that clumps form in
the wind, become dynamically un-coupled from the flow and line forces, and
if they have a certain range in clump mass, follow trajectories 
whereby they fall back toward the star. The bow shock around the clump would
account for the X-ray emission, and drag owing to the wind/clump velocity
difference would be important in the clump trajectory.

The idea that hot star winds are clumpy has been especially well
investigated in regards to Wolf-Rayet (WR) stars.  Lepine \& Moffat (1999)
found that emission features seen moving across broad optical lines
of \ion{He}{2} lines can be explained as clump emission regions moving
out in the winds. Hillier (1991) showed that unclumped winds of WR stars
would lead to line broadening by electron scattering that is not observed.
Nugis \& Lamers (2000) found that the observed IR flux distribution of
WR stars is better fit with clumped models.  There is also evidence for
clumping in the case of other early-type stars.  Lupie \& Nordsieck (1980)
found that the position angle of polarization seen in B supergiants shows
irregular changes, and they interpreted this as arising from very large
density enhancements in the winds. This was perhaps the first indication
of fragmentation at a {\it large} scale. Brown, Ignace, \& Cassinelli
(2000) showed that only large scale concentrations of matter can lead
to observable polarization changes, because smaller scale reshuffling of
electrons in an envelope does not lead to a significant net polarization
change.  In addition to the WR and B supergiants, there is 
new evidence of wind clumping in O star winds (Bouret \etal\ 2003;
Evans \etal\ 2004; Bouret, Lanz, \& Hillier 2005).

Clumping is important a) because it leads to overestimates of mass-loss rate
(\Mdot) values of early-type stars that are derived from density square
diagnostics (Abbott, Bieging, \& Churchwell 1981), and b) because the
downwardly revised \Mdot\ values affect the X-ray spectra owing to a reduced
absorption column density through overlying wind material. The best
estimates of \Mdot\ had long been assumed to be those derived from the
free-free radio flux of stars (Barlow \& Cohen 1976; Cassinelli \& Hartmann
1977).  The results were considered the most reliable because the free-free
transitions lead to a large (LTE) opacity; the observable radio flux tends
to form in the outer regions of the wind where the velocity is constant and
the density $\rho$ varies simply as $r^{-2}$; and the wind temperature
cancels from the emergent flux formula.

The topic of clumping has recently become a major one because surprisingly
large reductions to the mass-loss rates for O~stars have been suggested by
Massa \etal\ (2003) and Fullerton, Massa, \& Prinja (2006).  These authors
argue that a better estimator of \Mdot\ is the P-Cygni profile of the
\ion{P}{5} ($\lambda 1118, 1128 \Angs$) doublet.  This spectral feature has
several advantages: it originates in a dominant ion stage, so there is a
minimal ionization fraction correction and the profile is typically
unsaturated in UV spectra. Also, since the line opacity depends only
linearly on density, the line depth is not affected by clumping in the wind.
Fullerton \etal\ deduce mass-loss rate reductions by at least a factor of
1/10 to 1/100 times traditional values.  However, based on our clump
picture, we question the idea that the clumping does not affect the
abundance of the \ion{P}{5}, because the clumps are immediately adjacent to
X-ray sources, and thus the fraction of phosphorus in the observable
\ion{P}{5} stage could be reduced significantly owing to the multiple
ionizations associated with the Auger effect (Odegard \& Cassinelli 1982)

Another argument against the significant decrease in \Mdot\ is from WC07 who
find that the radii of formation of the X-ray lines corresponds quite well
with the values at which radial optical depth is unity when using the
traditional \Mdot\ values. Thus lines from soft ions such as that of \NVII\
(25 \Angs) form at 5 to 10 stellar radii because the opacity to radiation at
these wavelengths is large.  (Continuum opacity, much of it K-shell opacity
of abundant metals, varies roughly as $\lambda^3$.) Shorter wavelength
lines, that originate from the higher ions, such as \NeIX, \MgXI, \SiXIII,
can be seen as forming progressively deeper. These are also near their
respective optical depth unity locations, if traditional \Mdot\ values are
used.  Another good argument against the drastic decrease in the mass-loss
rates are that it would affect the well-established results of massive star
evolution (Hirschi 2007). The topic of clumping is an active one and at the
recent Potsdam international workshop on hot star wind clumps (Hamann,
Feldmeier, \& Oskinova 2007), a consensus was reached that \Mdot\ could not
be reduced by more than about a factor of 3 from the traditional values.

In summary, it has become important to understand the properties and the
effects of clumps in winds.  Our plan here is to consider the various
effects discrete clumps would have on observations of X-ray lines and to
establish an analytical tool to interpret observations. There is evidence
for clumps being important in astrophysics in general, for example, bow
shocks also appear in images of Herbig-Haro objects (e.g., Eisloffel \etal\
1994) and planetary nebulae (Odell \etal\ 1995). By using \Chandra\ data
we can obtain good information on the clumps in hot star winds. 

In \S 2 we summarize some of the problematic X-ray results obtained
from \Chandra.  In \S 3 we describe the numerical method we have used
and show the most relevant results regarding the post shock temperature
and density structure as well as the temperature distributed emission
measure. In \S 4, the bow shock structure is contrasted with plane
parallel shock pictures and in \S 5, we develop a simple ``on the
shock'' approximation to provide some insight regarding bow shocks.
In particular we give a derivation of the distributed emission measure
power law with temperature and discuss the angular flow around the clump.
Overall conclusions are summarized in \S6.

\section{Hot Star X-ray Problems}

At the forefront of current problems in this field is the finding
that X-ray line profiles from massive star winds are quite different
from what had been expected.  MacFarlane \etal\ (1991) had predicted
that lines formed by shocks in stellar winds should be blue shifted
and skewed. This is because the shock regions on the far side of the
star (where the gas is red shifted) would be more highly attenuated by
bound-free continuum opacity of the wind matter by virtue of being at
higher column density relative to the observer.  This effect was below
the spectral resolution of the emission lines of the two B stars (\ecma\
and \bcma) that were observable with the {\it Extreme Ultraviolet Explorer}
(\EUVE) satellite. These stars were observable at 70--730 \Angs\ and
500--700 \Angs, respectively (Cassinelli \etal\ 1995, 1996).  However,
well-resolved line profiles at the even shorter X-ray wavelengths are
observable with the high energy and medium energy grating spectrometers
(HEG and MEG) on \Chandra. The \Chandra\ observations (Waldron \&\
Cassinelli 2001; Miller \etal\ 2003; Leutenegger \etal\ 2006; WC07)
do not show the expected blue-ward skewing of the lines nor blue
shifting of the line centroids.  The lines tend also to be broad,
although much narrower for the low luminosity classes of hot stars,
and for all luminosity classes the HWHM is less than the terminal wind
speed. This problem of symmetry alone has motivated several ideas.

\begin{description}

\item A) If the mass-loss rates were reduced by an order of magnitude
or more, the winds would be sufficiently thin that radiation from both
the near {\it and} far side could escape (Cohen \etal\ 2006;
Leutenegger \etal\ 2006).

\item B) Owing to clumps, the winds could be more porous to the transfer
of X-rays (Oskinova \etal\ 2006).  The presence of high
density fragments allows the X-rays to escape from deeper in the wind,
and one could detect sufficient X-rays from the far side of the star to
eliminate the net shifting and skewing effects.

\item C) There is enhanced Sobolev escape of line radiation out the side
of shocks (Ignace \& Gayley 2002). This means that the line radiation
is coming in large part from the ``sides'' of the star, as seen by the
observer and this occurs where the radial flow  has a small line-of-sight
velocity.  Although we will not be employing the Sobolev escape argument,
we will find the enhanced sideways escape to be a useful concept.

\end{description}

In the study of clump effects by Feldmeier \etal\ (2003) and Oskinova
\etal\ (2004), the clump regions are pictured to be in the form of plane
parallel slabs. These correspond to regions of enhanced absorption of the
X-rays that originate in broader more diffuse wind regions. WC07 note that
the derived radii of the line source regions found by these authors are
larger than the rather small radii that WC07 find from an analysis of the
forbidden-intercombination-resonance ({\it fir}) triplet lines of He-like
ions. Oskinova \etal\ (2004) have developed plausible explanations for
the unshifted un-skewed lines that are observed in $\zeta$ Orionis. The
clumps in their view arise from a runaway effect first described in
Feldmeier \etal\ (1998). Whether porosity at the required level can
explain the observations has been questioned by Owocki \& Cohen (2007).
Nevertheless, the Oskinova \etal\ picture combines several elements:
the fragmentary nature of shocks and the possibility of producing X-rays
at a range of radii that is needed in any explanation of hot star X-rays.

\section{A Hydrodynamical Calculation of a Wind Colliding
with an Impenetrable Object}

We use the magnetohydrodynamical (MHD) code that was developed by Cho \&
Lazarian (2002) to model the bow shock that forms from a plane-parallel
flow impinging upon a spherical blunt obstacle. As a first step in
understanding the effects of bow shocks on hot star X-ray emission, we
choose to assume that the clump is impenetrable. Actual clumps are likely
to be more complicated, and the clumps could have a highly time-dependent
interaction with the incident wind.  However to ensure that we are
not dealing with numerical noise phenomena, we treat the simplest case
possible as a starting point for our theoretical exploration of bow shock
effects.  The main expectation is the formation of a bow shock around
the clump that exhibits a range in hot plasma temperatures along with a
non-trivial vector velocity flow.  This recipe is suggestively promising
for explaining the observed X-ray spectral features from hot stars.

For the simulation, the clump is assumed stationary and spherical
with a radius \Rc. The incident flow is plane-parallel at constant
speed and is parameterized by the mass flux and flow speed.  In such
an idealized case, the X-ray emission arises owing to the post-shock
gas that envelopes the clump. All of the emission arises {\it from
the wind matter} that is heated, compressed, and redirected at the bow
shock interface. For the purpose of interpreting X-ray observations,
we are interested in finding the temperature and density structure,
and the distributed emission measure. Using these physical properties
in conjunction with a cooling function provides the X-ray emissivity of
a blob as a function of wavelength and the line profile emission for a
particular transition of interest.

For computing the structure from bow shocks, we use a third-order
hybrid essentially non-oscillatory (ENO) MHD routine as described
in Cho \& Lazarian (2002) and references therein to reduce spurious
oscillations near shocks, two ENO schemes are combined. When variables
are sufficiently smooth, the 3rd-order Weighted ENO scheme (Jiang \& Wu
1999) is used without characteristic mode decomposition. When the opposite
is true, the 3rd-order Convex ENO scheme (Liu \& Osher 1998) is used. A
three-stage Runge-Kutta method is employed for time integration. The ideal
MHD equations are solved with magnetic field set to zero everywhere.
The simulation runs until it reaches an approximately steady-state bow
shock, usually after about two wind crossing times owing to the highly
hypersonic nature of the flow.  The simulations reported in this paper
are axisymmetric. We adopt cylindrical coordinates $(\varpi, \phi, z)$
with incident flow moving in the $+z$-direction and so the bow shock
is axisymmetric about the $z$-axis.  We shall frequently refer to the
transverse coordinate $\varpi$ as the impact parameter.  Some early
results were presented by Moeckel \etal\ (2002).

In the calculations a polytropic relation with $P \propto \rho^{\gamma}$ is
assumed so that the temperature structure is determined by the expansion
cooling of the gas after it passes through the bow shock geometry. 
Adiabatic cooling is an appropriate approximation for the case of high
velocity inflow, so the post-shock region is at a high temperature where
radiative losses are less important than expansion.  The computed results 
for the shape of the shock are quite similar to the semi-analytic results 
of Lomax \& Inouye (1963) for a $\gamma = 5/3$ bow shock.  

The incident velocity is specified as a Mach Number $M_W$, and here we
present the results of an analysis of an adiabatic bow shock with wind flow
$M_W=47$ and 71, that correspond to relative velocities of 1000 and 1500
\kms, respectively.  Throughout this paper, the subscript ``$W$'' refers to
wind properties.  A fast wind is assumed to be incident upon a rigid
stationary sphere, creating the bow shock structure.  The sphere has a
radius of 32 grid-points. The center of the sphere was located one half grid
spacing below the lowest grid-point. The results of this simulation are
robust for other large values of $M_W$ since the shock shape, flow pattern,
and shock structure are all nearly independent of Mach number in the
hypersonic limit of $M_W\gg 1$ for an adiabatic case (Hayes \& Probstein,
1966).  The parameters used in our simulations are given in
Table~\ref{tab:simulationparameters}.

The most basic parameters for determining the structure of a bow shock
around a blunt object are the incident velocity, which determines the
post-shock temperature, and the incident mass flux $\rho_W \Vrel$.
The numerical calculation is made in the rest frame of the rigid sphere
where $\Vrel$ is the difference between the radial wind velocity and
the radial clump velocity.  The post-shock gas temperature follows the
well known Rankine-Hugoniot relation between the velocity perpendicular
to the shock and the post-shock temperature.  The maximum temperature,
\TA\ achieved occurs at the ``apex'' of the bow shock (i.e., along the
line of symmetry) and is given by

\begin{eqnarray}
\TA  & =& \frac{3}{16}~ \frac{\mu m_H}{k}~( V_{\rm rel,\perp}^2) \\
       & =& 14~\mbox{MK} \, \left(\frac{V_{W,\perp}}{1000~\kms}\right)^2 
\label{eq:threesixteenths}
\end{eqnarray}
where in the latter expression we have evaluated the constants
using fully ionized solar abundances and
$mu=0.62$, and $V_{W,\perp}$ is the
perpendicular speed of the pre-shock gas relative to the shock front. 

%-----------------------------------------------------------------------

\begin{figure}

%\centering{\includegraphics[height=16cm, angle=0]{f1.eps}}
\plotone{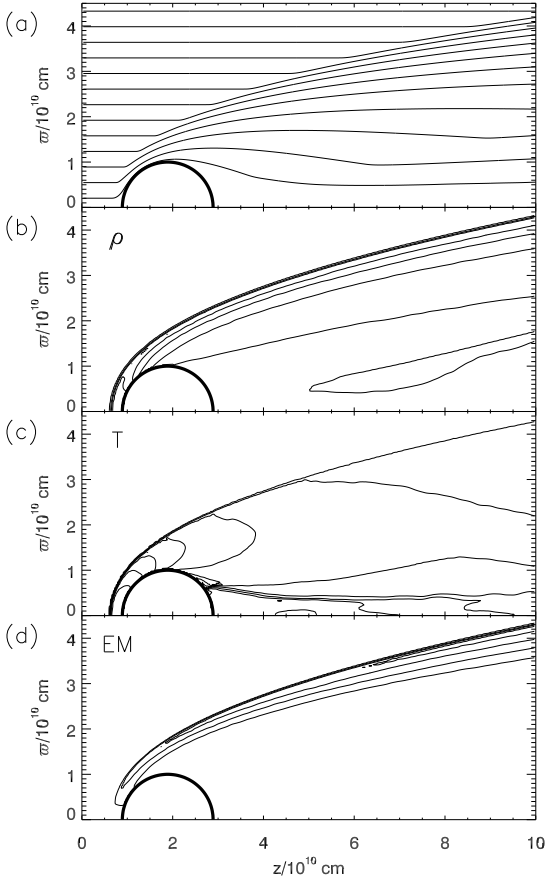}
\caption{
Simulation results of an adiabatic bow shock forming in
response to an impenetrable sphere using the parameters given in
Tab.~\ref{tab:simulationparameters} with $V_W=1500$ \kms.  Shown are (a)
flow streamlines, (b) density with contours at 0.4, 1.2, 2.0, 2.8, and
3.6$N_W$, (c) temperature with contours at 1, 3, 6, 10, 15, 21, and 28 MK,
(d) emission measure contours of $2\pi \varpi N^2$ separated by factors
of two with the largest value at $2.44\times 10^{22}$ cm$^{-5}$ and
showing that the emission measure is isolated strongly at the shock front.
}
\label{fig:4plotcontours}
\end{figure}

\begin{figure}
%\epsscale{0.05}
%\centering{\includegraphics[width=5in,angle=0]{f2.eps}}
\plotone{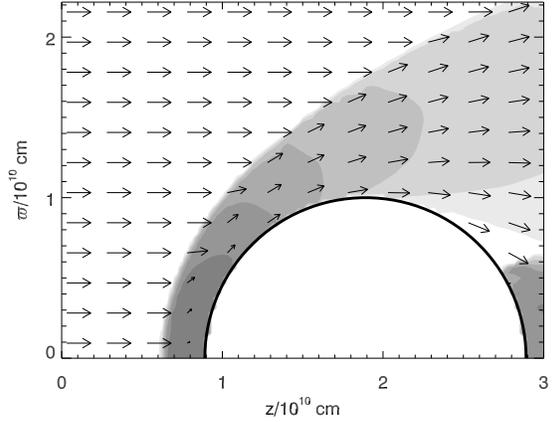}
\caption{A contour plot of temperature in the bow shock simulation.
The peak temperature contour value near the stagnation region is $\TA$
given by eq.~(\ref{eq:threesixteenths}), and the contours in temperature
have the same spacing as in Fig.~\ref{fig:4plotcontours}. Arrows show
the velocity vectors.}
\label{fig:Tbowvec}
\end{figure}

Figure \ref{fig:4plotcontours} shows simulation results from
our adiabatic bow shock case using the parameters listed in
Table~\ref{tab:simulationparameters} (using the second set of parameters
wherever two values are listed).  Figure~\ref{fig:4plotcontours}
shows (a) streamlines of the flow, (b) the density, (c) temperature,
and (d) emission measure distributions . Figure~\ref{fig:Tbowvec}
shows temperature contours behind the bow shock superposed with
vectors detailing the post-shock flow velocity.  Note especially the
decreasing temperature along the bow shock with distance from the apex,
and also that the flow of gas around the clump leads to a significant
transverse vector velocity field in conjunction with the temperature
distribution.  Note also, that all the X-ray emission properties, in
which we have special interest here, closely hug the bow shock surface.
This is what inspires our introduction of the ``on the shock'' or \OTSH\
approximation. In the following section, we consider observables relating
to these properties of the clump-wind simulation.

\section{The Temperature Distribution of the  Emission Measure}

\subsection{Results for a Single Clump}

An important property for understanding processes that generate hot
plasma in astrophysical sources is the temperature distribution of the
emission measure. This distribution  is often called the ``differential
emission measure'', but the use of this term in the X-ray literature is
not uniform, and we prefer simply to refer to the temperature distribution
of the emission measure or \EMT. The contribution to the emission measure
at a given temperature range from $T$ to $T+dT$ can arise from totally
disconnected regions in the X-ray emitting region.  The volume emission
measure (EM) is defined as

\begin{equation}
EM = \int \, N_{\rm e}\,N_{\rm p}\,dV,
\end{equation}
a volume integral involving the product of the electron and proton number
densities $N_{\rm e}$ and $N_{\rm p}$.  Wojdowski \& Schulz (2004, 2005;
hereafter WS04 and WS05) find that the amount of hot plasma at each
range in temperature to be a decreasing power law for most hot stars,
with the exception of those few hot stars known to possess highly
magnetized envelopes (e.g., \thoriC, see Donati \etal\ 2002).  Each of
the X-ray emission lines is associated with a range of temperatures
that can be found from the APEC software (Smith \etal\ 2001).  From the
strength of the line, Wojdowski \& Schulz 
find the interesting result
that \EMT\ is a downward sloping power law extending from about 2~MK, up
to an apparent maximum that is typically about 20~MK.  Such a distribution
holds for the majority of the stars that were analyzed.  For stars known
to be highly magnetic, the \EMT\ distribution can have a very different
shape. The highly magnetic star \thoriC\ even has a positive slope over
the whole temperature range indicating the presence of larger amounts
of increasingly hot material. Clearly such a distribution would have to
stop at some high temperature, although this is not seen in the \thoriC\
plots shown by WS05 or WS07. Table~\ref{tab:WojdSchulzEM} lists values
of the \EMT\ power law index for several OB stars as derived from the
analyses of WS05 and WC07.

\begin{figure}
%\centering{\includegraphics[height=12cm]{f3.eps}}
\plotone{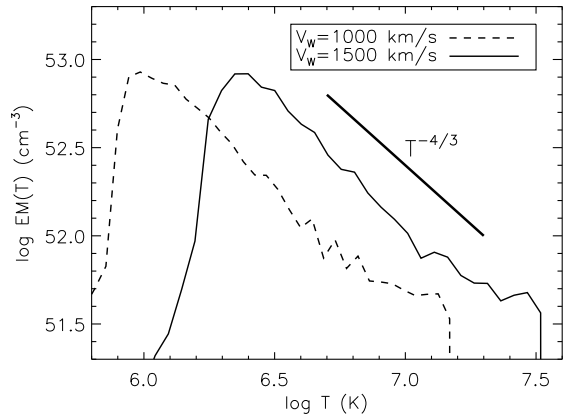}
   \caption{The
  temperature distribution of the emission measure from a simulation of an
  adiabatic bow shock using the parameters given in 
  Tab.~\ref{tab:simulationparameters}.  The maximum temperature \TA\ is set by the
  speed of the incident wind relative to the shock.  The low temperature
  cut-off is due to the finite size of the simulation grid.  The dashed line
  is a power law approximation with a slope of $-4/3$. Two results are
  shown: these have $M_W =47$ and 71, which correspond to the \Vrel = 1000
  \kms and \Vrel =1500 \kms as indicated} 
\label{fig:bowEM}
\end{figure}

Using the hydrodynamical code, we obtain a power-law emission measure
distribution $EM(T) \approx 10^{51.4}\;{\rm cm}^{-3} \times (T/\TA)^p$,
shown in Figure~\ref{fig:bowEM}, where $p$ is approximately $-4/3$.
Note that the drop-off of at low temperatures is an artifact of the
calculation that arises from the fact that the grid in the impact
parameter direction is truncated at some $y_{\rm max}$. When this maximal
impact parameter is increased the slope continues to follow the $-4/3$
slope to lower temperature values.  With this power-law distribution,
the temperature gradient $dEM/ dT \propto T^{p-1} \propto T^{-7/3}$,
a result we will use later on.  This figure was obtained by binning
the emission measure results of each simulation volume element into
temperature cells of $0.1$ in $\log T$.

It is useful to have the following scaling of the results of the
simulations to parameters for a different clump:  with radius $R$,
relative wind/clump velocity $V_W$, wind density $N_W$ ($=\sqrt{N_{\rm e}
N_{\rm p}}$), and apex temperature \TA.  We derive the scaling

\begin{eqnarray}
\frac{d EM}{d T} &= &\frac{10^{51.4} 
{\rm cm}^{-3}}{14~{\rm MK}}\, V_8^{-2}\, \nonumber \\
 & &
\left(\frac{R}{R_{\rm s}}\right)^3 \left(\frac{V}{V_{\rm s}}\right)^2
\left(\frac{N}{N_{\rm s}}\right)^2 \left(\frac{\TA}{T}\right)^{4/3}
\left(\frac{\Delta \log T}{0.1}\right)
\end{eqnarray}
where $V_8$ is the incident speed in $10^8$~cm/s. Here we have used
subscript, $s$, for our simulation results.  The $p=-4/3$ slope in 
Figure~\ref{fig:bowEM} somewhat underestimates the emission from the highest
temperature bin in the distribution (by a factor of about 2), and we
discuss this later in our analytic derivation of the $dEM/dT$ relation.

From an interpretational perspective, the most important feature to
note about the power law is that there is significantly more gas at low
temperatures than at the high temperatures near the apex of the shock.
The increasing emission measure for the lower temperatures (and hence the
lower ion stages) arises from the fact that a bow shock has an increased
area toward the wings where the shock is increasingly oblique.
As a result of the power law distribution \EMT, clumps that are deep in a 
wind can produce a great deal of low ion emission, say at \OVII.  In the
case of stars with high mass-loss rates, such as Of~stars and 
OB~supergiants, the wind opacity can substantially block this emission from
direct observations. 

The stars that have been studied most intensively in the history of wind
theory and hot star X-ray astronomy are \zori\ and \zpup, and both have
\EMT\ slopes reasonably close to the $p = -4/3$ value.  However some stars,
such as \tsco, have a very shallow but negative slope. It has recently
been discovered that this particular star is also one with a strong
magnetic field (Donati \etal\ 2006), thus it is likely that this star
has a combination of X-rays from both the clump regions and the positive
sloping \EMT\ magnetic regions. Sorting the contributions from each could
be difficult; however, it appears that the \EMT\ analyses can
yield valuable 
new information about clump properties and may also provide an indicator
of fields in stars that have not yet shown measurable Zeeman effects. We
have found a case of a positive sloping \EMT\ in our recent calculations
of X-rays from Be stars (Li \etal\ 2007) using a Magnetically Torqued Disk
model for Be stars. This model is based on the idea that the matter that
enters a Be star disk is from a wind that has been channeled by a dipolar
magnetic field and which is co-rotating with the star (Cassinelli \etal\
2002; Brown \etal\ 2004). Higher temperature gas originates farther from
the star since the impact of the wind with the disk is at a higher speed.
In this example there is a maximum to the \EMT\ distribution that is
directly related to the magnetic field at the base of the wind.
Next we consider the consequences of having a large number of clumps
contributing to the \EMT.

\subsection{Ensemble of Clumps}

An inspection of  the results from WS05 reveals that, except for the
magnetically dominated star \thoriC, observed \EMT\ distributions have
negative power-law slopes.  However the values are not exactly the
$-4/3$ that we derive from the clump simulation.  This can arise for
several reasons. The first one has already been discussed, namely some
stars are strongly magnetic and those fields can dominate the \EMT\
following physics that is quite distinct from clump bowshocks, and so
the results are not surprisingly quite different.  However the star
HD 206267A (O~6.5V(f)) has a slope that is steeper than the $-4/3$
value. In WC07 we found that for near main-sequence stars,
one could detect source emission coming from a wide range in depths as
the winds are not as optically thick as those in the supergiants. Thus
a second way to explain a deviation from a $-4/3$ slope in the case of
steeper \EMT\ is to consider the X-rays as arising from an {\it ensemble}
of different clumps.

Such an ensemble does not represent only more or fewer clump bowshocks
but in fact a range of \TA\ values.  Each individual clump contributes a
\EMT\ distribution of slope $-4/3$ for $T < \TA$, but now clumps exist
at different radii and thus a range in \TA\ may plausibly exist from a
range in \Vrel\ values.  The consequence of staggered values in \TA\ leads
naturally to a slope that is steeper than $-4/3$.  All clumps contribute
to low temperature plasma, but only a small minority contribute to the
absolute maximum temperature achieved in the entire wind.  Note that
one way to provide an observational cut-off for lower temperatures
is to recognize that even rather low mass-loss winds will eventually
lead to substantial photoabsorption at sufficiently low X-ray energies.
Second, our simulation assumes the clump is small compared to the radius
of the star, such that the wind is plane parallel on the scale of the
clump size.  However, the stellar wind is in fact spherically divergent,
and so the flow striking the a large bow shock is actually more oblique
than is achieved in the plane-parallel flow simulation.

\subsection{Adiabatic and Radiative Cooling Regions}

Before leaving the topic of \EMT, it is important to comment on the
applicability of the adiabatic assumption.  Across a bow shock there
is a large range in temperatures, and even if the adiabatic assumption
holds near the apex of the shock, it will fail somewhere out in the
wings, since radiative cooling is extremely efficient for $T < 10^6 K$.
So it is relevant to consider the regimes in which the cooling time is
small or large compared to the flow time in the simulation.  

Adiabatic cooling is a result of expansion cooling as the flow navigates
around the clump.  Throughout this trajectory the gas is emitting X-rays,
and so the gas is cooling by radiation as well; however, the adiabatic
assumption adopted in our hydrodynamical simulation will apply to those
cases when the radiative cooling timescale is slow relative to expansion
cooling.

To characterize the comparison, we define $t_{\rm flow}$ as the flow time
scale associated with adiabatic cooling.  This will scale as $t_{\rm
flow} \approx \Rc/\Vrel$.  With $\Rc = 10^{10}$~cm and a wind speed of
500~\kms\ (approximately a quarter of the terminal speed for a typical
O star wind), the flow time works out to $t_{\rm flow} \sim 10^3$~s.
We define the radiative cooling time with $t_{\rm rad}$, which can
be estimated as the ratio of the thermal energy density $U_{\rm th}$
to the cooling rate $dU/dt$.  Thus, $t_{\rm rad} \approx U_{\rm th} /
(dU/dt) = (N_{\rm e} kT) / [\Lambda (T) N_{\rm e}^2]$, where $\Lambda$
is the cooling function in erg cm$^3$ s$^{-1}$ which is tabulated in Cox
(1999), giving the result $\log \Lambda= -21.6 - 0.6\, (\log T - 5.5)$.
At a value of $T=10^7$~K associated with \TA\ and a number density of
$N_{\rm e} =10^{10}$~cm$^{-3}$, the radiative cooling time becomes $t_{\rm
rad} \sim 10^4$~s, about ten times larger than the flow timescale under
these conditions.

Clearly, the radiative cooling is dominant over adiabatic cooling where
the wind density is large or the shock temperature is low, and in future
simulations it will be important to include radiative cooling.  However,
it is still useful to consider the the limiting case of adiabatic cooling.
First, although 1D simulations show radiative cooling to be important,
expansion cooling is there limited to $r^{-2}$ divergence, and shock
structures are necessarily spherical shells in these simulations, which we
know not to be true based on low levels of X-ray variability (Cassinelli
\& Swank 1983).  Our 2D simulations allow for a new geometric avenue
of expansion cooling because of the channeling of the gas around the
clump in the form of a bowshock, thus requiring a new assessment of how
adiabatic cooling contributes to the interpretation of hot star X-rays.
Second, in the hypersonic limit, the results for pure adiabatic cooling
are quite robust because the bowshock shape and temperature distribution
are independent of density.  This is not the case for radiative cooling,
hence one expects the radiative bowshock properties to be different
for clumps at different radii in the wind.  And third, as discussed in
the preceding section, we can explore the extent to which this limiting
case can reproduce observed differential emission measure distributions.
Departures from the model predictions are suggestive of contributions
by radiative cooling or other effects, such as stellar magnetism.

\section{The ``On the Shock'' Approximation}

From the numerical modeling, we find, in contrast with planar
shocks (e.g., Feldmeier \etal\ 1997), that the X-ray emission is
dominated by a zone lying just behind the shock front (as seen in
Fig.~\ref{fig:4plotcontours}d). Hence, we introduce the ``On the Shock''
or \OTSH\ approximation for a simple analysis of bow shocks. In the
case of a planar frontal shocks, the post-shock flow must continue in
a straight line, and thus the only cooling that occurs is radiative
emission.  In the case of a bow shock, there is always an expansion of
the gas around the sides of the clump and is associated with adiabatic
cooling.

Consider a simple but useful picture for the shock structure.  As before,
let the $z$ axis be the radial direction from the star through the apex of
the shock and through the center of the spherical clump.  Let $\varpi$ be the
perpendicular direction that corresponds to the impact parameter of the
wind flow relative to the center of the clump.  The shape of the bow shock
from our simulation can be fit with a power-law curve given by

\begin{equation}
\frac{z - z_\circ}{\Rc} = a\left(\frac{\varpi}{\Rc}\right)^{m},
\label{eq:zGynumbs}
\end{equation}
with $a=0.35$ and $m=2.34$, hence a shape not far from a parabola.
This provides a good fit out to $\varpi \sim 5$ clump radii. An exact
solution to the shock shape derived using an inverse method is given in
Lomax \& Inouye (1963), and our result for the shock shape agrees well
with theirs.  Their paper does not provide the temperature and emission
measure information of special interest to us. In addition to the shape,
it will also be useful to know the derivative of the shape (i.e., the
position-dependent tangent).  We define this to be $g(\varpi) = \tan
\Ain = dz/d\varpi$ where \Ain\ is the angle that the incident wind
makes relative to the bow shock.

\subsection{Post-shock velocity components}

\begin{figure}
%\centering{\includegraphics[height = 10cm, angle=0]{f4.eps}}
\plotone{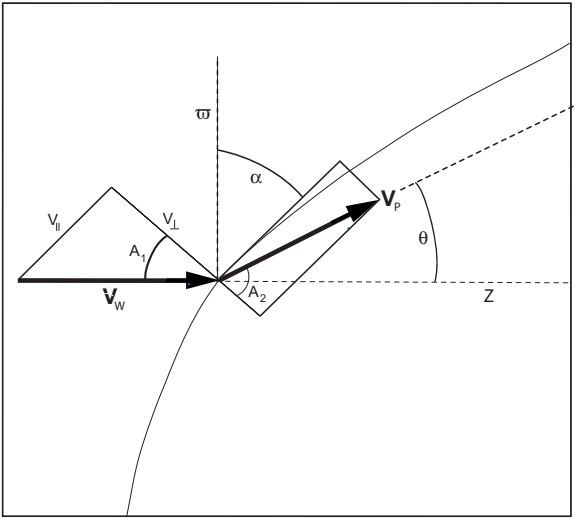}
\caption{An illustration of the changes in velocity components at the
  shock front.  The angles $A_1$ and $A_2$ are the incident and
  emergent angles of the flow relative to the direction perpendicular
  to the bow shock, and $\theta$ represents the angle between
  the post-shock velocity vector and the $z$ direction.  The impact
  parameter of a streamline is given by its $\varpi$ coordinate value.}  
\label{fig:velfig}
\end{figure}

Figure \ref{fig:velfig} shows a post-shock (hereafter identified with
subscript ``P'') flow trajectory associated with crossing the bow
shock. The jump conditions for the velocity components perpendicular
and parallel to the shock front are:

\begin{eqnarray}
\VPperp &=& \frac{1}{4} \VWperp\\ 
\VPpara &=& \VWpara 
\label{eq:VPperpar}
\end{eqnarray}
All across the face of the shock the incident wind speed is given by $
\Vvec = V_W \zvec$.  Based on the geometry of Figure~\ref{fig:velfig} 
we can derive the following relations:
\begin{equation}
\VWperp =  \VW ~\cos (\Ain) \hspace{.2cm} =  \VW ~\frac{1}{\sqrt{1+g^2}} 
\label{eq:VWper}
\end{equation}
\begin{equation}
\VWpara   =  V_W~ \sin (\Ain) \hspace{.2cm} =  V_W~ \frac{g}{\sqrt{1+g^2}} 
\label{eq:VWpar}
\end{equation}
where we have expressed the velocity quantities in terms of the shape
gradient $g(\varpi)$.  The following limiting behavior is implied: as $g \ll 1$,
the wind is essentially normal to the shock as occurs along the stagnation
line, and at large impact parameter where $g\gg 1$, the shock becomes
nearly parallel to the incident wind flow.

Consider a jump that occurs at some position on the shock ($\varpi,
z$).  The incident wind speed \VW\ relative to the orientation of the 
shock front can be expressed as $\VW^2 = \VWperp^2 +\VWpara^2$.
The total post-shock velocity is $\VP^2
= (\VWperp/4)^2 + \VWpara^2$, which upon using substition
using the relations (\ref{eq:VWper}) and (\ref{eq:VWpar}) becomes

\begin{equation}
\VP^2 = \VW^2 ~~ \frac{[\frac{1}{16} + h_1^2]}{[1+h_1^2]},
\label{eq:VPsq}
\end{equation}
where on the post-shock side, the velocity vector makes an angle relative to
the normal implicitly defined by
$h_1 \equiv \tan \Aex= \VPperp/\VPpara= 4 \,\tan
\Ain$. Thus,

\begin{eqnarray}
\VPperp &=&  \VP ~\cos \Aex  = \VP ~ \frac{1}{\sqrt{1+h_1^2}} \\
\VPpara &=&  \VP ~\sin \Aex  = \VP ~ \frac{h_1}{\sqrt{1+h_1^2}} 
\label{eq:VPperpar_h1}
\end{eqnarray}

\subsection{Temperature distribution across the bow shock}

\begin{figure}
%\centering{\includegraphics[height=12cm]{f5.eps}}
\plotone{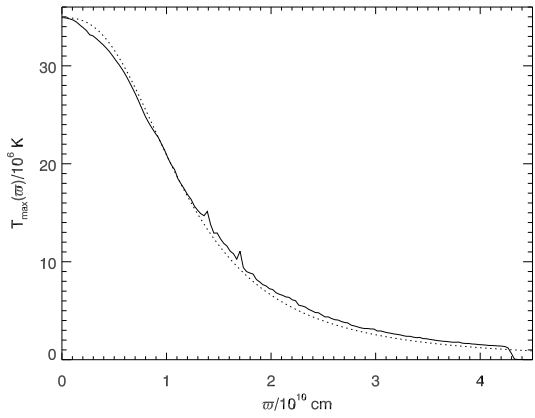}
\caption{The maximum temperature as a function of impact parameter in
  our simulation of an adiabatic bow shock.  Shown are simulation
  results (solid) and eq.~\ref{eq:temp_otsh} (dotted).} 
\label{fig:temp_vs_otsh}
\end{figure}

The post-shock temperature as a function of
impact parameter $\varpi$ is related to the
change in the perpendicular velocity component as described in
the previous section. The temperature
distribution is given by

\begin{eqnarray}
T_P(\varpi) &=& \frac{3}{16} \frac{\mu m_H}{k} \,
           (\Vvec~ {\mathbf \cdot} ~ \nvec)^2\\ 
     &=& \TA ~\cos^{2}(\Ain) \nonumber \\  
     &=& \frac{\TA}{1+g^2(\varpi)} 
\label{eq:TP}
\end{eqnarray}
where \TA\ is the highest temperature gas, at the bow shock apex.
Thus we see that each impact parameter point on the shock has
a specific post-shock temperature associated with it, 
hence formally $\varpi = f(T)$. The simple power law for the bow shock geometry
in (\ref{eq:zGynumbs}) leads to 

\begin{equation}
\frac{dz}{d\varpi} \equiv g(\varpi) = a\, m\, (\varpi/\Rc)^{m-1},
\end{equation}
thus, we get an explicit equation for temperature in the \OTSH\ approach:

\begin{equation}
\frac{T}{\TA} = [1~+~g^2]^{-1} = \left[~1~ + ~0.67\left( \frac{\varpi}{\Rc} \right)^{2.68}
\right]^{-1}
\label{eq:temp_otsh}
\end{equation}
This expression is plotted against the maximum temperature along an
impact parameter in Figure~\ref{fig:temp_vs_otsh} where the agreement
between the model and the preceding expression is remarkably good.
This close agreement is related to the conditions of the simulation
being hypersonic.

\subsection{Polar angle distributions of the X-rays}

To facilitate the use of these results for evaluating observables from
an ensemble of clumps, such as in predicting emergent line widths,
it is convenient to express the angles of the inflow and post-shock
flow relative to the local ``radial'' (or $z$) direction. This is
a blob-centered $(r,\theta,\phi)$ system.  The incident radial wind
flow has $\theta_{in} = 0^\circ$. On the post-shock side, the angle is
$\theta_P = \Aex - \Ain$, hence the stream flow emerges from the bow shock at

\begin{equation}
\tan \theta_P = \frac{3g}{1+4g^2} \equiv h_2
\label{eq:tanthP}
\end{equation}
Thus the ``transverse'' or sine component of the post-shock velocity vector
is $\sin \theta_P=h_2/\sqrt{ 1+h_2^2}$.

What is particularly interesting about the post-shock velocity and its
direction is that it affects the observed line profiles.  The post-shock
speed can, for example, have a rather large transverse value, as 
illustrated in Figure~\ref{fig:Tbowvec}. To choose one case for example,
consider the one angle quadrature value of Lucy \& White (1981): $\Ain =
30^\circ$ then $\Aex = 67^\circ$, $\VP = 0.56 \VW$ and the transverse
component of the post-shock velocity is one-third of $\VW$. This is already a
significant fraction of observed X-ray line widths (in WC07 this was typically
about $0.4 \VW$). In a simulation involving many clumps, one would
need to account for the line-of-sight velocity of each of the clumps. 
The narrower lines found by WC07 for the near main sequence stars 
and for the lower ion stages can be explained by having the line formation 
regions being closer to the star where the local wind speed is smaller than 
is the case for the supergiants. For the latter case the radiation must
arise from regions above about optical depth unity, and this is well out
in the wind where the wind speed is near  the fast terminal speed.

The transverse component of the X-ray emitting matter is another
significant difference from planar shocks.  The planar
shocks are usually assumed to be perpendicular to the flow, and thus the
post-shock matter continues flowing only perpendicular to the shock front. 

\subsection{Temperature distributions of the emission measure}

Semi-analytic expressions for bow shock properties have been derived before,
especially for the simple case that we consider here in which the compact
object is an impenetrable sphere.  For example Canto \& Raga (1998) derived
the shock shape and pressure density structure for an isothermal bow shock.
This is valid for cases in which the cooling time is rapid and so the shock
material forms a thin sheet around the blunt object. In contrast our specific
interest is in the hot adiabatic region where X-ray emission originates.

Our results for the post-shock density have motivated our
``on the shock'' approximation that provides a temperature, velocity,
and $EM$ from each spatial element of the bow shock surface.  These are
useful for understanding some of the outstanding questions raised
by high spectral resolution X-ray data (\eg\ Burke \etal\ 2006).
In a separate paper, the \OTSH\ approximation will be employed in
quantitative spectral synthesis to model X-ray line profile shapes.
Here, we present a derivation for $EM(T)$ to show that the power-law
temperature distribution arises naturally from the bow shock geometry.

Based on our hydrodynamic simulation, the OTSh approximation asserts
that the hot gas contributing to $EM$ closely hugs the bow shock surface.
In this region the density is nearly constant at $N=4N_W$ everywhere
along the bow shock.  Hence, the volume element, $dV$, associated with
$dEM$ must scale roughly as the incremental surface area $dS$ of the
bow shock times some depth parameter which we call $\DL$.  This $\DL$
parameter represents a characteristic depth that is perpendicular
to the bow shock surface. 
Then our expression for $dEM$ is given by

\begin{equation} 
 dEM = N^2 \, dS \, \DL
	\label{eq:dEM}
\end{equation}
The differential surface area for any function $\varpi = f(z)$ used for
a surface of revolution about the z-axis is given by 

\begin{equation} 
 dS = 2 \pi \, \varpi\, \left[ 1 + \left(\frac{d\varpi}{dz}\right)^2 \right]^{1/2} \,dz
\end{equation}
Recalling that $g=dz/d\varpi = $, $T/\TA = (1+g^2)^{-1}$, and defining $\tau
= T/\TA$, the rate change in surface area with the monotonic temperature
``coordinate'' along the surface becomes

\begin{eqnarray}
\frac{dS}{dT} & = & 2\pi\,\varpi\, \left( 1 + \frac{1}{g^2}\right)^{1/2} \, \frac{dz}{dT}\nonumber \\ 
 & = & 2\pi\,\varpi\, \frac{\sqrt{1+g^2}}{g}\,\frac{dz}{d\varpi}\,\frac{d\varpi}{dT}\nonumber \\ 
 & = & \tau^{-1/2}\,\left( 2\pi\,\varpi\,  \frac{d\varpi}{d\tau} \right)\nonumber\\ 
 & = & \tau^{-1/2}\,\frac{d}{d\tau} (\pi \varpi^2).
\end{eqnarray}
This rather elegant result requires yet one more step to fully
determine the surface gradient entirely in terms of temperature;
that step requires the solution to $\varpi(T)$.  The mapping between $\varpi$
and $T$ is made through the factor $g$ using equations~(\ref{eq:zGynumbs})
and (\ref{eq:temp_otsh}), yielding the implicit relation of

\begin{equation}
g = am\,\left(\frac{\varpi}{\Rc}\right)^{m-1} = \sqrt{\frac{1}{\tau}-1}.
\end{equation}
Solving for $\varpi$, taking the gradient with temperature, and doing
some algebraic manipulation yields finally

\begin{eqnarray}
\frac{dS}{d\tau} & = & -\pi\,\Rc^2\,(am)^{-2/(m-1)} \,\tau^{-(3m-1)/(2m-2)} \nonumber \\
 & & 	\left( 1-\tau\right)^{-(m-2)/(m-1)}. 
	\label{eq:dSdtau}
\end{eqnarray}

We know that the result of the simulation for the emission measure is
$dEM/dT \propto T^{-7/3}$, and we know that the value of $m$ for the bow
shock is only a little steeper than a parabola.  Using equations
(\ref{eq:dEM}) and (\ref{eq:dSdtau}), it is convenient therefore
to express the final result for the emission measure gradient as

\begin{equation} 
\frac {dEM} {dT} = - \frac{\pi\,N^2\,\Rc^2\,\DL} {(am)^{2/(m-1)}\,\TA} \,
	\tau^{-2.5+\alpha}\,\left(1-\tau\right)^{-\alpha},
\label{eq:demdtfin}
\end{equation}
where $\alpha = (m-2)/(m-1)$.  If the bow shock were exactly a
parabola, then $m=2$, $\alpha=0$, and $dEM/dT \propto T^{-2.5}$
is a pure power law with an exponent remarkably close to the value
of $-7/3 = -2.33$.  With $m=2.34$ and $\alpha = 0.25$, the OTSh
approximation predicts an emission measure gradient of $dEM/dT
\propto T^{-2.25}\,(1-\tau)^{-0.25}$. In the limit that $T \ll \TA$,
we have that $\tau \ll 1$ and $dEM/dT$ is again a power law with an exponent
that is now even closer to the fit value.  Obtaining an exact
match to the fit exponent of the simulation requires $m = 2.20$
in the low $\tau$ limit, not far from the $m=2.34$ value
derived from the bow shock shape.  It would appear then that the OTSh
is an excellent approximation for the emission measure and temperature
distributions of the bow shock in the hypersonic and adiabatic limits.

There is one problem with equation~(\ref{eq:demdtfin}) in that near the
bow shock apex, $\tau$ approaches unity in which case $dEM/dT$ becomes
singular for $\alpha >0$.  This singularity is however only a mathematical
artifact.  The cusp arises from our fit to the $z(\varpi)$ formula with a
power-law exponent of 2.34, but in fact the analytic result of Lomax
\& Inoye indicates that very near the apex the curve is a parabola in
which case $dEM/dT$ is non-singular.  Earlier we noted that the $p=-4/3$
law underestimates the $EM$ very close to the apex temperature.  It is
interesting that the parameterization of equation~(\ref{eq:demdtfin})
does lead to a rise in $EM(T)$ when $\tau$ is not negligible.  In practice
the singularity can be avoided by applying a high temperature cut-off,
in which case the OTSh provides an excellent prescription for $z(\varpi)$,
$T(\varpi)$, and $EM(T)$.

With regards to the $\DL$ factor, based on the simulations, achieving
the good results for $dEM/dT$ in terms of $S(T)$ means that
the range of constant density along the bow shock is consistent
with $\DL$ a constant width.  We find a value of $\DL \approx 0.1
\Rc$, independent of the  $y$ location along the bow shock.
It is now useful to define an emission measure
scaling parameter $EM_\circ$, using $a=0.35$ and $m=2.34$, giving 

\begin{eqnarray} 
EM_{\circ} & = & \frac {\pi\,N^2\,\Rc^2\,\DL} {(am)^{2/(m-1)}} \nonumber \\
 & = & 6.8 \times 10^{51} 
	~ {\rm cm}^{-3}\, \left( \frac {\Rc} {10^{10}} \right)^3 \left(\frac
{N_W} {10^{10}}\right)^2 \left( \frac {\DL} {R_C}\right),
\end{eqnarray}
where we assume a strong shock such that $N = 4 N_W$, and \Rc\ and
$N_W$ have both been scaled to the values used in the simulation.
Using $\DL = 0.1 \Rc$, the scale constant becomes $EM_\circ \approx
7\times 10^{50}$~cm$^{-3}$.  As shown by WC07, typical $EM$ values for
line formation are $\approx 10^{55}$~cm$^{-3}$, implying that about
$10^4$ clumps would be required.  Assuming simple Poisson statistics,
the expected variability would then be about 1\%, consistent with
the low levels of X-ray variability from OB~stars that has been previously
noted.  Interesting with $EM_\circ \propto \Rc^3$, even small changes
in the clump radius -- certainly a poorly known quantity -- has a
significant impact on the predicted numbers of the clumps and expected
levels of variability suggesting that X-ray variability is a useful
means of constraining clump properties (e.g., Oskinova \etal\ 2001).

Although the giants and MS stars with lower mass-loss rates (smaller optical
depths) show more dispersion in the range of $R_{fir}$, there is no evidence
of any X-ray emission arising from below the associated optical depth unity
radii.  Furthermore, regardless of luminosity class, none of the stars show
high energy ion stages forming far from the star. In the clump bow shock
picture this means that the relative speed associated with the shocks is
well below the ambient wind speed, which would produce very high
temperatures.  Clumps far from the star are likely being dragged out by the
wind, and this reduces the shock temperatures in the outer regions, but with
a sufficient shock jump in speed to produce the relatively low ion stages
such as \OVII, and \NVII. 

As for the observed near-zero velocity shift and broad HWHM of the lines,
these can best be explained if we have the X-ray line emission arising
from the sides of the star as seen by the observer (i.e., perpendicular to
the line-of-sight). This region contains a spread in speeds around the,
$v_z=0$, iso-velocity surface. Such a concentration of emission from
that sector of the wind would occur if there were ``self absorption''
occurring. That is, each cool clump tends to absorb the line emission
originating at the bow shock at the star-ward side of the clump. Thus we
see an enhanced contribution from the regions at, and symmetric about,
$v_z = 0$. This counters the expected tendency to see the blueward-shifted
and -skewed line emission.

\section{Summary}

One of the advantages of the clump bow shock picture that it allows for
a consideration of a 3-dimensional model for the X-ray production from
OB star winds. The bow shocks can be distributed randomly about the star
or in some other pre-specified way. Similarly one could choose to assume
a variety of clump sizes at any radial shell. So a variety of ideas can
be tested with the bow shock picture.  Accounting for bow shocks around
clumps in winds can potentially explain a number of properties of the
X-ray emission observed from hot stars.

{\it The wide range of ionization stages} that occurs at essentially every
radial distance from the star can be produced by bow shocks because of the
power law distribution of $EM(T)$.  Low ion stages such as \OVII\ occur far
from the star, but not the highest ions.  One might think that the shocks
would be strongest there because the wind speed is the largest. However,
clumps could be dragged out by the wind and thus there could be a lower
relative speed at large r, as discussed by Howk \etal\ (2000).

{\it The zero centroid shift problem, $( V_{peak} \approx 0.0)$} 
can be explained if the line radiation is generated primarily from clumps that are
to the side of the star. This could occur because of
the ``self-absorption'' effect, whereby the  clumps along the line-of-sight to 
the star strongly attenuate the X-rays  that are produced on the  
starward side of the clumps. 

{\it The broad line widths or HWHM of up to about half the wind speed} can
be caused by a combination of the sideways velocity the bow shock flow plus
a range in $V_z$ values for an ensemble of clumps.  This is in contrast
with planar shocks that have been proposed for the shock fragments in
hot star winds.  The transverse velocity in the bow shocks is an essential
aspect of the adiabatic or expansion cooling of the bow shock gas.

{\it The narrower line of main sequence and giant stars} can be explained
by the correlation of wind speed with the transverse velocity around a
clump. These stars have optically thin winds and the emission can be
dominated by shocks located deeper in the wind where the inflow speeds are
slower.

{\it The lines of high ion stages are not narrow} as one might expect from
source regions close to the star. However, to a large part this can be
explained by the relatively low spectral (or velocity) resolution. For
the \Chandra\ HETGS/MEG spectra the HWHM resolution is about 270 \kms
at 25 \Angs\ for lines of low ion stages, and to 1300 \kms\ for the high
ions. The transverse velocity behind a clump and the range in clump
velocities would lead to further broadening.

{\it There are observed \EMT\ power laws (\eg\ WS05),} and these result
directly from the wind-oblique shock interactions. The shape of the \EMT\
distribution depends on the fraction of the wind material lying above
optical depth unity.

%{\it The concentration of the  highest ions and highest temperatures
%relatively near the star} could arise if clumps can have large speeds
%relative to the wind near the star and this could arise from clump
%trajectories or perhaps by plasmoid ejections from the base of the wind.

{\it The $R_{fir}$  radius corresponds rather well to the radius of optical
depth unity,} and this  can probably be
explained by having the observed line formation sources located to the sides of
the star relative to the line-of-sight to the observer and having
the mass loss rate reduced by an acceptable amount such as 1/3
the traditional \Mdot\ values.

In addition to a consideration of the problems summarized above, we have
introduced the \OTSH\ picture. This provides a way to derive the line source
information and the line of sight Doppler velocity shifts needed to compute line
profiles from a single clump or an ensemble of clumps 
distributed through a wind.

The clump bow shock picture forms a useful element in a more realistic
3-D view of the stellar winds of hot stars, and the \OTSH\ approach
forms a useful technique for predicting and analyzing X-ray emission
properties. The results should be of interest to areas other than hot
star astronomy.  Clump flow interactions are thought to occur in such
objects as RS Oph (Nelson \etal\ 2008). Clumps with bow shock interfaces
with a wind are seen in Hubble images of planetary nebulae (Odell \etal\
1995).  An understanding of clumps can also affect 
research well beyond the stellar wind community.

\acknowledgments We thank Nick Moeckel for early work on the $EM(T)$
relation and Greg Tracy for producing some of the figures.  We are also
grateful to John Brown for many interesting conversations on the topic
of wind clumps.  JPC, RI, AL and NAM have been supported in part by
award TM3-4001 issued by the \Chandra\ X-ray Observatory Center. We also
acknowledge support from the NSF Center for Magnetic Self Organization in
Laboratory and Astrophysical Plasmas.  WLW acknowledges support by award
GO2-3027A issued by the \Chandra\ X-ray Observatory Center.  \Chandra\
is operated by the Smithsonian Astrophysical Observatory under NASA
contract NAS8-03060.

%%%%%%%%%%%%%%%%%%%%%%%
%%%%%%%%%%%%%%%%%%%%%%%
%% Start of Tables and Figures
%%%%%%%%%%%%%%%%%%%%%%%

\begin{deluxetable}{cc}
\tablewidth{4.0in}
\tablecaption{Bow shock simulation parameters.}
\startdata
\hline\hline
Simulation Parameter & Value(s) \\
\hline
$R_c$          &  $10^{10}$ cm                                   \\
$(Mach\#)_W$   &  47 \& $71$                                     \\
$V_W$          &  $1000~\&~1500~\mbox{km~s}^{-1}$                \\
$N_{W,e}$      &  $10^{10}$ cm$^{-3}$                            \\
%$\mu$          &  $0.62$                                         \\
$T_W$          &  $2   \times 10^4~\mbox{K}$                     \\
$T_S$          &  $1.4\times 10^7 \&\; 3.2\times 10^7~\mbox{K}$    \\
$\rho_W V_W^2$ &  100 \& 233~ dyne cm$^{-2}$
\enddata
\label{tab:simulationparameters}
\end{deluxetable}

%%%%%%%%%%%%%%%%%%%%%%%%%%%%%%
%%%%%%% TABLE 2
%%%%%%%%%%%%%%%%%%%%%%%%%%%%%%

\begin{deluxetable}{lcc}
\tablewidth{4.0in}
\tablecaption{ Emission Measure Slopes versus Temperature} 
\startdata
\hline\hline
Star & Spectral Type & Slope $p$   \\
     &               & {\small $( = 1+d\log EM/d\log T)$}   \\      

\hline
\zpup               & O4f                  & $-$1.1 \\
\zori               & O9.7 Ib              & $-$1.2 \\
\iori               & O9 III               & $-$1.3 \\
\dori               & O9.5II               & $-$0.9 \\
HD206267            & O6.5 V               & $-$2.2 \\
$\beta$ Cru         & B0.5 III             & $-$2.1 \\
$\tau$  Sco         & B0.2 V               & $-$0.6 \\
$\theta^1$ Ori C    & O4-6p                & $+$1.5
\enddata
\label{tab:WojdSchulzEM}
\end{deluxetable}

%%%%%%%%%%%%%%%%%%%%%%%%%%%%%%
%%%%%%% END TABLES
%%%%%%%%%%%%%%%%%%%%%%%%%%%%%%

%%%%%%%%%%%%%%%%%%%%%%%%%%%%%%
%%%%%%% START FIGURES
%%%%%%%%%%%%%%%%%%%%%%%%%%%%%%

%%%%%%%%%%%%%%%%%%%%%%%%%%%%%%
%%%%%%% FIGURE 1 murphbowshock4plot.eps dec15 2007
%%%%%%%%%%%%%%%%%%%%%%%%%%%%%%

\newpage

%Simulation results of an adiabatic bow shock forming in
%response to an impenetrable sphere using the parameters given in
%Tab.~\ref{tab:simulationparameters}.  Shown are (a) flow streamlines,
%(b) density with contours at 0.4, 1.2, 2.0, 2.8, and 3.6$N_W$, (c)
%temperature with contours at 0.1, 0.2, 0.3, 0.5, 0.7, and 0.9 \TA, (d)
%emission measure contours of $2\pi \varpi N^2$ separated by factors of
%two with the largest value at $2.44\times 10^{22}$ cm$^{-5}$ and showing
%that the emission measure is isolated strongly at the shock front.

%%%%%%%%%%%%%%%%%%%%%%%%%%%%%%
%%%%%%% FIGURE 2  Tbowvec
%%%%%%%%%%%%%%%%%%%%%%%%%%%%%%

\newpage

%%%%%%%%%%%%%%%%%%%%%%%%%%%%%%
%%%%%%% FIGURE 3 bowEM = murphy15dec
%%%%%%%%%%%%%%%%%%%%%%%%%%%%%%

\newpage

%%%%%%%%%%%%%%%%%%%%%%%%%%%%%%
%%%%%%% FIGURE 4 Velfig
%%%%%%%%%%%%%%%%%%%%%%%%%%%%%%

\newpage

%%%%%%%%%%%%%%%%%%%%%%%%%%%%%%
%%%%%%% FIGURE 5 Compare Ts
%%%%%%%%%%%%%%%%%%%%%%%%%%%%%%

\newpage

\end{document}